\documentclass[12pt]{article}
\usepackage{amssymb, latexsym, amsmath, hyperref}
\usepackage{epsfig}
\usepackage{graphicx}
\usepackage{amsfonts}
\usepackage{mathrsfs}
\usepackage{multirow}
\usepackage[dvips]{color}

\newcommand{\be}{\begin{equation}}
\newcommand{\ee}{\end{equation}}
\newcommand{\bea}{\begin{eqnarray}}
\newcommand{\eea}{\end{eqnarray}}

\newcommand{\ed}{\end{document}}

\newcommand{\bi}{\begin{itemize}}
\newcommand{\ei}{\end{itemize}}

\newcommand{\bce}{\begin{center}}
\newcommand{\ece}{\end{center}}

\begin{document}

\title{Spectral Singularities Do Not Correspond to Bound States in the Continuum}

\author{Ali~Mostafazadeh\thanks{E-mail address:
amostafazadeh@ku.edu.tr, Phone: +90 212 338 1462, Fax: +90 212 338
1559}
\\
Department of Mathematics, Ko\c{c} University,\\
34450 Sar{\i}yer, Istanbul, Turkey}

\date{ }
\maketitle

\begin{abstract}
We show that, contrary to a claim made in arXiv:1011.0645, the von~Neumann-Winger bound states that lie in the continuum of the scattering states are fundamentally different from Naimark's spectral singularities.
\vspace{2mm}


\end{abstract}

In 1927 von~Neumann and Wigner constructed a spherically symmetric scattering potential that supported a bound state with a positive energy \cite{VNW}. This corresponded to a genuine, square integrable solution of the time-independent Schr\"odinger equation.  Because the continuous spectrum of the Schr\"odinger operator coincided with the set of nonnegative real numbers, this bound state was called ``a bound state in the continuum.'' This class of bound states was subsequently  studied by various authors, notably Stillinger and Herrick \cite{SH}, Gazdy \cite{gazdy}, and Friedrich and Wintgen \cite{FW} who interpreted them as resonances with a zero width.

In 2009 the present author has revealed the physical meaning and possible applications of the mathematical concept of a spectral singularity \cite{prl-2009}. This is a generic feature of complex scattering potentials that was discovered by Naimark in the 1950's \cite{Naimark} and studied thoroughly by mathematicians for more than half a century \cite{math}. Because spectral singularities correspond to poles of the reflection and transmition coefficients (and of the S-matrix) and belong to the continuous spectrum of the Schr\"odinger operator, which is real and nonnegative, they can also be interpreted as zero-width resonances \cite{prl-2009}. This has led to the claim \cite{rotter} that states associated with spectral singularities are nothing but the bound states in the continuum.

A quick look at the properties of spectral singularities show that they correspond to scattering solutions of the  time-independent Schr\"odinger equation. In particular, they are not square-integrable. This is in contrast with bound states in the continuum that are associated with square-integrable solutions of the  time-independent Schr\"odinger equation. The following are other major differences between spectral singularities and bound states in the continuum.
    \begin{enumerate}
    \item A real scattering potential can never support a spectral singularity \cite{math}. This is certainly not the case for bound states in the continuum. In fact, the scattering potentials introduced by von~Neumann and Wigner  \cite{VNW} and others \cite{SH,gazdy} that admit a bound state in the continuum are real.
    
    \item Spectral singularities appear for generic complex scattering potentials, while scattering potentials that involve bound states in the continuum are extremely rare. For example, scattering potentials with a compact support cannot have a bound state in the continuum, whereas a complex potential with a compact support can easily admit a spectral singularity \cite{prl-2009,pras}.
    \end{enumerate}
    
The above discussion shows that although both the bound states in the continuum and spectral singularities can be viewed as zero-width resonances, they are quite different in nature. Most notably, spectral singularities have a much wider domain of application, because they occur for generic complex scattering potentials. A concrete evidence is provided by the observation that every lasing system involves a spectral singularity. This corresponds to lasing at the threshold gain \cite{pra-2011a}.
    
    \vspace{.3cm}
\noindent {\em \textbf{Acknowledgments:}} This work has been supported by the Turkish Academy of Sciences (T\"UBA).

\end{document}